Can we obtain the predictive value of GP-B experiment from the known experimental results? This predictive value is more reliable then that deduced from special model of theory. In this paper, we calculate in this way. The result is same as that from special relativistic gravitational theory. So it is extremely likely that GP-B experiment will prove space-time is flat.
//

**(1) The conclusion deduced from the results of three classical relativistic experiments**

The perihelion shift of planet and deflection of light are the problem of a body moves in the gravitational field due to a static sphere $M_0$. The experimental result of the perihelion shift of planet is

$$\Delta\varphi^{(e)} = \frac{6\pi GM_0}{c^2 a(1-e^2)} \tag{1}$$

And the experimental value of deflective angle of light is

$$\boldsymbol{d} = \frac{4GM_0}{c^2 R} \tag{2}$$

The experimental value of variance of frequency is

$$\boldsymbol{n} - \boldsymbol{n}_0 = \frac{-GM_0 \boldsymbol{n}}{c^2 R} \tag{3}$$

according to the red shift experiment. The red shift of light is also the problem of a photon moves in the gravitational field in substance. Using the Einstein's relation between energy and frequency $w = h\boldsymbol{n}$ and (3), we obtain that the variance of energy is

$$w - w_0 = \frac{-GM_0 w}{c^2 R} \tag{4}$$

if a body moves from $r_0 = R$ (the sun) to $r \approx \infty$ (the earth). (4) is the another form of experimental result from red shift.

**What can we learn from these experimental results?**

From Newton gravity we obtain that, if a body moves in the field due to a static sphere, its variance of energy $\Delta w$ and variance of angular momentum $\Delta L$ are

$$\Delta w = GM_0 m \left(\frac{1}{r} - \frac{1}{r_0}\right) = \frac{GM_0 w}{c^2}\left(\frac{1}{r} - \frac{1}{r_0}\right) \tag{5}$$

$$\Delta L = 0 \tag{6}$$

respectively. Comparing (4) and (5), we obtain that, if a body moves in the field due to the static source, only the Newton gravity contributes to the variance of energy, but the set of (5) and (6) can not explain the perihelion shift of the planet and the deflection of light. So the only correct way is to suppose that, the gravity contains an additional term except the Newton' term

$$\mathbf{F} = \mathbf{F}^{(N)} + \mathbf{F}^{(A)} \tag{7}$$

There are the characters of the additional force:

- From the experiment of red shift, only the Newton gravity is to do work, the additional force is not to do work, so the direction of the additional force perpendicular to the velocity $\mathbf{u}$ of the exerted body.

- The direction of normal line of the orbital plane $\mathbf{r}^0 \times \mathbf{u}$ is invariance in the procedure of motion of the planet or photon. However the direction of Newton gravity is perpendicular to $\mathbf{r}^0 \times \mathbf{u}$, so the direction of the additional gravitational force is also perpendicular to $\mathbf{r}^0 \times \mathbf{u}$. Summing up above two points, the direction of additional gravitational force is in the direction $\mathbf{u} \times (\mathbf{r}^0 \times \mathbf{u})$.

- If the body moves in the gravitational field due to a static sphere, analogy to the Newton gravity, the additional gravity must be proportional to the gravitational constant $G$, the mass of static sphere $M_0$, the mass of the exerted body $m$, then the additional gravity is

$$\mathbf{F}^{(A)} \propto GM_0 m \mathbf{u} \times (\mathbf{r}^0 \times \mathbf{u}) \tag{8}$$

- All of the gravitational field strength satisfies same equation, so there must be the same factor of $r$ in the expression of each component of field. It means that the additional gravity must contain $r^{-2}$ factor in the static case. In addition, the additional gravity has the same dimension as Newton gravity, therefore

$$\mathbf{F}^{(A)} = \frac{\boldsymbol{l}\, GM_0 m}{c^2 r^2} \mathbf{u} \times (\mathbf{r}^0 \times \mathbf{u}) \tag{9}$$

where $\boldsymbol{l}$ is a quantity with non-dimension, it determines the strength of the additional gravitational force relative to the Newton gravity. Therefore when a body is in the field due to a static sphere, the force exerted on it is

$$\mathbf{F} = \frac{GM_0 m}{r^3}\left[-\mathbf{r} + \frac{\boldsymbol{l}}{c^2}\mathbf{u}\times(\mathbf{r}\times\mathbf{u})\right] \tag{10}$$

Now we introduce the mathematical construction of the gravitational force from the experimental results. There is only a coefficient $\boldsymbol{l}$ waiting to determine in (10). From this formula, we obtain

$$\frac{d\mathbf{L}}{dt} = \mathbf{r}\times\mathbf{F} = \frac{-\boldsymbol{l}\, GM_0 m}{c^2 r^2} u_r (\mathbf{r}\times\mathbf{u}) = \frac{\boldsymbol{l}\, GM_0 \mathbf{L}}{c^2}\frac{d(1/r)}{dt} \tag{11}$$

Therefore we have

$$L = L_0 \exp\left[\frac{\boldsymbol{l}\, GM_0}{c^2}\left(\frac{1}{r} - \frac{1}{r_0}\right)\right]$$

$$\Delta L = \frac{\mathbf{l}}{c^2} \frac{GM_0 L_0}{r} \left( \frac{1}{r} - \frac{1}{r_0} \right) \tag{12}$$

It means that, we must use (12) to substitute (6).

Now we discuss the perihelion shift of planet. From (5) and (12), we obtain

$$u^2 = c^2 \left[ 1 - k^2 \exp\left( \frac{-2GM_0}{c^2 r} \right) \right] \tag{13}$$

$$r^2 \frac{d\mathbf{j}}{dt} = h \exp\left[ \frac{(\mathbf{l}-1)GM_0}{c^2 r} \right] \tag{14}$$

Therefore we have

$$\left( \frac{d\mathbf{r}}{d\mathbf{j}} \right)^2 = -\left\{ 1 - 2[(\mathbf{l}-1)^2 - \mathbf{l}^2 k^2] \frac{G^2 M_0^2}{c^2 h^2} \right\} \mathbf{r}^2 + \\ 2[(1-\mathbf{l}) + \mathbf{l} k^2] \frac{GM_0}{h^2} \mathbf{r} + (1-k^2) \frac{c^2}{h^2} \tag{15}$$

where $\mathbf{r} = 1/r$. The solution of this equation has the form

$$\mathbf{r} = A[1 + e\cos(\mathbf{m}\mathbf{j})] \tag{16}$$

Substituting it into (15), we get

$$\mathbf{m} = 1 - \frac{(1-2\mathbf{l})GM_0}{c^2 a(1-e^2)} \tag{17}$$

$$\Delta\varphi^{(t)} = \frac{(1-2\lambda)2\pi\ GM_0}{c^2 a(1-e^2)} \tag{18}$$

Comparing (18) with (1), we obtain $\mathbf{l} = -1$.

We can make an analogous calculation for the deflection of light. Comparing the theoretical value with the experimental value, we obtain similarly $\mathbf{l} = -1$. We get the same $\mathbf{l}$ value from the motions of planet and photon, their velocities are extreme different. This fact means that, (10) is reasonable. It is correct that the size of additional gravity is proportional to the second power of velocity, except a $m$ factor. Substituting the value of $\mathbf{l}$ into (10), we get

$$\mathbf{F} = \frac{GM_0 m}{r^3}[-\mathbf{r} - \frac{1}{c^2}\mathbf{u}\times(\mathbf{r}\times\mathbf{u})] = m\left( \mathbf{E} + \frac{1}{c^2}\mathbf{u}\times\mathfrak{R}\cdot\mathbf{u} \right) \tag{19}$$

where

$$E_i = \frac{-GM_0 x_i}{r^3}$$

$$R_{il} = \frac{GM_0}{r^3} \begin{pmatrix} 0 & x_3 & -x_2 \\ -x_3 & 0 & x_1 \\ x_2 & -x_1 & 0 \end{pmatrix} \qquad (20)$$

Rewrite (19) as 4-dimensional form

$$K_i = m_0 \boldsymbol{g}^2 \left( E_i + \frac{1}{c^2} R_{kl} u_j u_l \right) = m_0 \left( H_{ijl} U_j U_l + H_{i44} U_4 U_4 \right) \qquad (21)$$

$$K_4 = \frac{i\boldsymbol{g}}{c} F_i u_i = \frac{i\boldsymbol{g}}{c} F_i^{(N)} u_i = m_0 \frac{E_i}{c^2} U_i U_4 = m_0 (H_{4i4} + H_{44i}) U_i U_4 \qquad (22)$$

where $\boldsymbol{g} = 1/\sqrt{1-\boldsymbol{b}^2}$. This set of equation is deduced in the case that the source is static. According to the demand of special relativity, the general form of gravitational force must be

$$K_n = m_0 H_{nsr} U_s U_r \qquad (23)$$

Therefore we obtain that the 4dimensional gravity is the quadratic function of 4dimeensional velocity. This formula of gravity and the strengths of field are deduced from the experimental results, there are apart from the specific theoretic model of gravity. However this formula of gravity is same as the special relativistic gravitational theory or general relativity. From (23), we can introduce an additional condition

$$H_{nsr} = H_{nrs} \qquad (24)$$

Comparing (23) with (21) and (22), and using (24), we obtain that the strengths of gravitational field due to a static sphere are

$$H_{ill} = \frac{-GM_0 x_i}{c^2 r^3} \qquad \text{does not sum for } l \qquad i \neq l$$

$$H_{lil} = H_{lli} = \frac{GM_0 x_i}{2c^2 r^3} \qquad \text{does not sum for } l \qquad i \neq l$$

$$H_{i44} = \frac{GM_0 x_i}{c^2 r^3}$$

$$H_{4i4} = H_{44i} = \frac{-GM_0 x_i}{2c^2 r^3} \qquad (25)$$

$$H_{111} = H_{222} = H_{333} = H_{444} = 0, \qquad H_{il4} = H_{i4l} = H_{4il} = 0$$

$$H_{123} = H_{231} = H_{312} = H_{132} = H_{321} = H_{213} = 0$$

**(2) The strength of gravitational field due to a moving sphere and the precession of gyroscope of orbital effect**

Suppose that the instantaneous velocity of satellite is **v** and along the direction of $ox_1$ observed in the earth static frame, then in the satellite instantaneous static frame, the strength of gravitational field is

$$H_{nsr}' = a_{nx} a_{sh} a_{rz} H_{xhz} \tag{26}$$

where

$$a_{nx} = \begin{pmatrix} g & 0 & 0 & i\beta g \\ 0 & 1 & 0 & 0 \\ 0 & 0 & 1 & 0 \\ -i\beta g & 0 & 0 & g \end{pmatrix} \tag{27}$$

Substituting (25) into (26), we may find out all of $H_{nsr}'$. The components of strength field relative to the precession of gyroscope is

$$H_{ij4}' = H_{i4j}' = \frac{i\beta\gamma GM_0}{c^2 r^3} \begin{pmatrix} x_1/2 & -\gamma x_2 & -\gamma x_3 \\ 2\gamma x_2 & -x_1/2 & 0 \\ 2\gamma x_3 & 0 & -x_1/2 \end{pmatrix} \tag{28}$$

where $x_i$ and $r$ are the quantities measured in the earth static frame.

Let

$$B_i = ic^2 (H_{jk4}' - H_{kj4}') \tag{29}$$

$$P_{ij} = ic^2 (H_{ij4}' + H_{ji4}') \tag{30}$$

We can deduce the formula of angular acceleration of gyroscope from (23).

$$\frac{d\mathbf{w}}{dt} = \frac{1}{2}\left[\mathbf{w}\times \mathbf{B} - \mathbf{w}\cdot \wp + \mathbf{w} Sp(\wp)\right] \tag{31}$$

For the orbiting gyroscope, because the satellite is in the accelerated motion, so we must add a Thomas precession term, then the total angular acceleration is

$$\frac{d\mathbf{w}}{dt} = \frac{1}{2c}\left[\mathbf{w}\times \mathbf{B} - \mathbf{w}\cdot \wp + \mathbf{w} Sp(\wp)\right] - \frac{GM_0 (\mathbf{r}\times \mathbf{v})\times \mathbf{w}}{2c^2 r^3} \tag{32}$$

where the second term is the Thomas procession term.

Substituting (28)-(30) into (31), then take the average value for a circle, finally we obtain that the angular acceleration of the orbital effect is

$$\frac{d\mathbf{w}}{dt} = \frac{GM_0}{c^2 r^3}(\mathbf{r}\times \mathbf{v})\times \mathbf{w} \tag{33}$$

It is just the corresponding one predicted by the special relativistic gravitational theory. The angular acceleration of geodetic effect predicted by general relativity is

$$\frac{d\mathbf{w}^{(G)}}{dt} = \frac{3GM_0}{2c^2 r^3}(\mathbf{r}\times \mathbf{v})\times \mathbf{w}^{(G)} \tag{34}$$

Then we may say, if the Lorentz transformation is correct, the experiment of gravity probe-B will obtain the precession of gyroscope of orbital effect predicted by the special relativistic gravitational theory. If people obtain our predicted value from the GP-B experiment, then we can prove that the explanation of the perihelion shift of planet and the deflection of light given by general relativity are wrong.

REFERENCES

Zhang Junhao and Chen Xiang (1990). International Journal of Theoretical Physics, 29,579
Zhang Junhao and Chen Xiang (1991). International Journal of Theoretical Physics, 30,1091.
Zhang Junhao and Chen Xiang (1993). International Journal of Theoretical Physics, 32,609.
Zhang Junhao and Chen Xiang (1995). International Journal of Theoretical Physics, 34,429.
Zhang Junhao and Chen Xiang, Introduction to Special Relativistic Gravitational Theory.
Schiff,L.I.(1960). Proceedings of the National Academy of Sciences of the USA,46,871.